\documentclass{article}

\usepackage{arxiv}
\usepackage{graphicx}
\usepackage{cite}
\usepackage{mdframed}
\usepackage{empheq,mathtools}
\usepackage{amsmath,systeme}
\usepackage{mathrsfs}
\graphicspath{ {./images/}}
\usepackage[utf8]{inputenc} 
\usepackage[T1]{fontenc}    
\usepackage{hyperref}       
\usepackage{url}            
\usepackage{booktabs}       
\usepackage{amsfonts}       
\usepackage{nicefrac}       
\usepackage{microtype}      
\usepackage{lipsum}
\usepackage[nodayofweek,level]{datetime}

\title{A time-dependent SEIR model to analyse the evolution of the Sars-CoV-2 epidemic outbreak in Portugal}

\author{
  Pedro Teles \\
  Departamento de Física e Astronomia\\
  Faculdade de Ciências da Universidade do Porto\\
  Rua do Campo Alegre s/n, 4169-007 Porto \\
  \texttt{ppteles@fc.up.pt} \\
}

\begin{document}
\maketitle
\begin{abstract}
\textbf{ Background:}  The analysis of the Sars-CoV-2 epidemic is of paramount importance to understand the dynamics of the coronavirus spread. This can help health and government authorities take the appropriate measures and implement suitable politics aimed at fighting and preventing it.\\
\textbf{Methods:} A time-dependent dynamic SIR model inspired in a model previously used during the MERS outbreak in South Korea was used to analyse the time trajectories of active and hospitalized cases in Portugal. \\
\textbf{Results:} The time evolution of the virus spread in the country was adequately modelled. The model has changeable parameters every five days since the onset of mitigation measures. A peak of about 19,000 active cases is estimated.  Hospitalized cases could reach a peak of about 1,250 cases, of which 200-300 in ICU units.\\
\textbf{Conclusion:} With appropriate measures, the number of active cases in Portugal can be controlled at about 19,000 people, of which about 1,250 hospitalized and 200-300 in ICU units. This seems manageable by the country’s national health service with an estimated 1,140 ventilators.

\end{abstract}


\section{Introduction.}
There is already abundant information on Sars-CoV-2\cite{1,2,3,4,5}. In the most severe cases, the virus Sars-CoV-2 infection can lead to the development of acute respiratory distress syndrome (ARDS) causing respiratory failure, septic shock, multiorgan failure, and even death \cite{6}. Studies suggest that the case fatality rate of the virus is of about 3.5\% in mainland China \cite{7}. However, this value seems to be much higher in Italy \cite{8}, suggesting its strong dependence on demographics \cite{9}. The WHO declared Europe as the new epicentre of the disease on the 13th of March of 2020[\cite{10}. The rapid growth of the number of active cases presenting severe symptoms has saturated the health services in most countries in the continent, especially in Italy \cite{11}. Governments throughout Europe have implemented severe measures to prevent and mitigate the spread of the virus. Yet, as of the 6th of April there were as many as 646,340 confirmed cases on the European continent, resulting in 49,227 deaths\cite{2,3}. 

In this study, a time-dependent SEIR model \cite{12,13} was used to analyse the evolution of current active and hospitalised cases in Portugal. The model takes into account mitigation and self-protective measures implemented by the government and the population, from the 18th of March 2020 onwards. The use of time-dependent models has been proposed before as they provide a chance to readjust the parameters as time passes and the conditions in which the epidemic is spreading change\cite{14}.

This study shows that although sometimes blunt\cite{15}, an SEIR model can accurately predict the trajectory of the curve of infected and hospitalized cases, and may be suitable to be used in the future for a consistent analysis of the data, and the repercussions of mitigation and control measures, which can be taken into account in these models.

\section{Methodology.}

\begin{figure}
  \centering
  \includegraphics[width=\textwidth,height=\textheight,keepaspectratio]{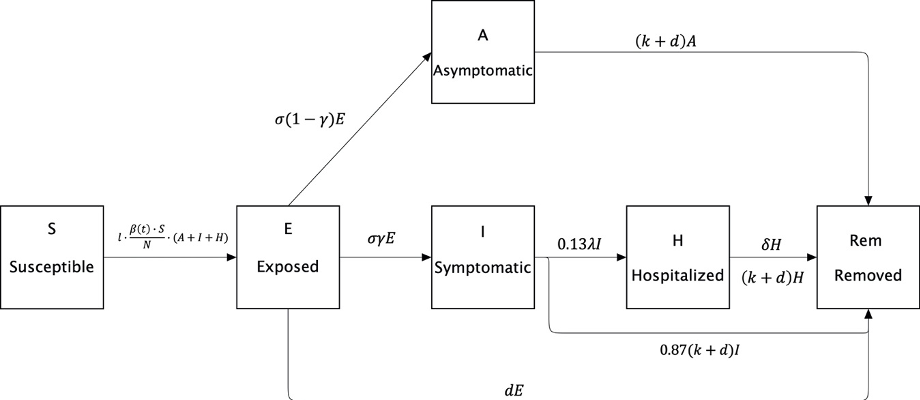}
  \caption{Flow chart of the SEIR model used in this work.}
  \label{fig:fig1}
\end{figure}

A time-dependent SEIR model inspired by a model developed by Xia et al\cite{12,13} was used, which can be understood by the flow diagram shown in figure \ref{fig:fig1}. This is an update from a previously used preliminary version [13], in which all infected cases ended up hospitalized. Given how unrealistic that sounds, the model now considers that 13\% of infected cases are hospitalized, and 87\% are recovered through the normal recovery rate (these proportions were taken from Portuguese data \cite{16}. Furthermore, the model considers only one transmission coefficient, one government measure coefficient, and one self-protective measure coefficient.

 In this model, S corresponds to the number of susceptible individuals; E, the number of exposed individuals; A, the number of asymptomatic cases; I, the number of mild-to-severe infected cases; H, the number of hospitalized cases; R, the number of removed cases, and finally N, the total population of Portugal.
 
 The model is time-dependent, because $\beta (t)$, the transmission coefficient of the asymptomatic, symptomatic, and hospitalized cases to the susceptible, can vary with time. Although this variation is not continuous. This will be explained later.

$\sigma^{-1}$ is the mean incubation period, $\lambda^{-1}$  is the mean time between symptom onset to hospitalization, $k^{-1}$  is the mean infectious/recovery period, $\delta^{-1}$  is the mean time from hospitalization to death, $\gamma$  is the clinical outbreak rate, $l$ represents the self-protective measures taken by individuals, and $d$  the mitigation measurements taken by the government. Of the symptomatic (I) cases 13\% need hospitalization, whereas 87\% heal without the need to be hospitalized.  The time unit is 1 day.
The set of differential equations can then be written as:


\begin{equation}
\systeme{
\frac{\mathrm{d} S(t)}{\mathrm{d} t} =  -l\cdot \beta (t) \cdot (\frac{S(t)\cdot A(t)}{N} + \frac{S(t)\cdot I(t)}{N}+ \frac{S(t)\cdot H(t)}{N}), 
\frac{\mathrm{d} E(t)}{\mathrm{d} t} =   l\cdot \beta (t) \cdot (\frac{S(t)\cdot A(t)}{N} + \frac{S(t)\cdot I(t)}{N}+ \frac{S(t)\cdot H(t)}{N}) -(\sigma+d)\cdot E(t), 
\frac{\mathrm{d} A(t)}{\mathrm{d} t} =  (1-\gamma)\cdot \sigma\cdot E(t)-(k+d)\cdot A(t), 
\frac{\mathrm{d} I(t)}{\mathrm{d} t} = \gamma\cdot \sigma\cdot E(t)-0.13\lambda\cdot I(t)-0.87(k+d)\cdot I(t), 
\frac{\mathrm{d} H(t)}{\mathrm{d} t} =  0.13\lambda\cdot I(t)-k\cdot H(t)-(\delta+d)\cdot H(t), 
\frac{\mathrm{d} R(t)}{\mathrm{d} t} =  k\cdot [A(t)+H(t)+0.87I(t)]+d[A(t)+I(t)+H(t)+E(t)]+\delta\cdot H(t)
}
\end{equation}

The only parameter considered to be known was the mean incubation time, which was taken as $\sigma^{-1}=5.1$ days from the literature (here using the median as equal to the mean) \cite{17}. From the previous work, determined from the Italian recovery data \cite{18}, a mean infectious/recovery period of about $k^{-1}\approx $ 11  days was used, and taken to be the same in all cases. The clinical outbreak rate which was taken to be $\gamma \approx$ 7\%, a mean value from values reported for attack rates \cite{19}, in a first approximation, and again, this is an approximate value. The epidemic starts at the 15th of February (the day of the first reported case in Europe) with only one person infected ($I_0=1$), being the rest of the values 0, except for $S_0$ which is simply N-1. See table \ref{table1} for a breakdown of the different parameters of this model.
To solve this system of differential equations the Mathematica code \cite{20} was used, using the function “NonLinearModelFit”\cite{21}. 

\begin{table}
\caption{The different parameters of the model described by Eq. 1}
\label{table1}
\centering
\begin{tabular}{lll}
\toprule
parameter      & Previous work \cite{13}  & This work  \\ \hline
$\sigma^{-1}$  & 5.1  (days)              & 5.1   (days)           \\
$\lambda^{-1}$ & $\sim$4 (days)            & To be fitted     \\
$\delta^{-1}$  & $\sim$59 (days)            & To be fitted     \\
$k_1^{-1}$     & 1/14         &   0.088 (now k)               \\
$k_2^{-1}$     & 0.088             (days$^{-1}$)                & 0.088 (now k)   \\
$\beta$        & 1.16$\pm$0.033) (days$^{-1}$)     & To be fitted (time-dependent)  \\
$\gamma$       & 0.0158$\pm$0.034) (days$^{-1}$)     & 0.07    \\
l              & Time-dependent    & To be fitted (time-dependent)     \\
d              & Time-dependent     & To be fitted  (time-dependent)    \\
N              & 10,290,000                      & 10,290,000       \\
$S_0$          & 10,289,999                      & 10,289,999       \\
$E_0$          & 0                             & 0                \\
$A_0$          & 0                            & 0                \\
$I_0$          & 1                              & 1                \\
$H_0$          & 0                              & 0                \\
$R_0$          & 0                              & 0                \\
\end{tabular}
\end{table}

In order to determine the parameters $\lambda$, and $\delta$, the number of hospitalized cases and deaths, respectively, as reported by the Portuguese Directorate-General of Health (DGS in Portuguese) and available online, were used \cite{16}. Fits were made on the 27th of March.
After determining the values for these parameters, they were used to fit the curve of infected cases in Portugal, where the values of $\beta(t)$, $d$ and $l$ were fitted in seven different periods:
\begin{itemize}
\item Period 0, from the 15th of February to the 18th of March (5 days after the government decreed a State of Emergency, giving the virus another incubation period),
\item Period 1, from the 19th to the 24th of March,
\item Period 2, from the 25th to the 30th of March,
\item Period 3, from the 31st of March to the 4th of April,
\item Period 4, from the 5th to the 10th of April,
\item Period 5, from the 10th to the 15th of April,
\item Period 6, from the 15th to the 20th of April (prediction).
\end{itemize}

The three parameters $\beta$, $l$, and $d$ were fitted for each of these five periods. All other parameters were fixed.

\section{Results.}
\subsection{ Adjusting parameter $\lambda$.}

\begin{figure}
  \centering
  \includegraphics[width=\textwidth,height=\textheight,keepaspectratio]{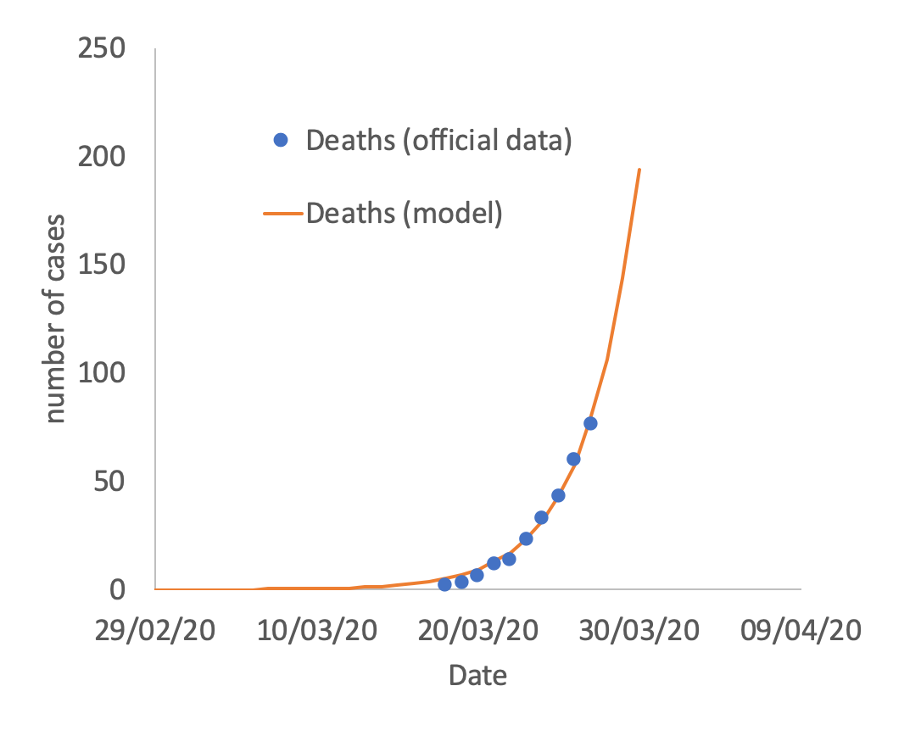}
  \caption{Graphical representation of the adjusted model to the Portuguese official death toll.}
  \label{fig:fig2}
\end{figure}

The number of deaths in Portugal was taken from \cite{16}, and the set of differential equations described in eq. 1 fitted to obtain the best possible fit for $\delta$. An $R^2$ value of $\sim$0.99 was obtained with the fit. The rest of the parameters are presented in table \ref{table2} . Results are shown in figure \ref{fig:fig2}.
\begin{table}
\caption{Table of fitted parameters  $\lambda$ and $\delta$ as obtained using “NonLinearCurveFit” in Mathematica}
\label{table2}
\centering
\begin{tabular}{lllll}
\toprule
parameter & estimate & standard error & t-statistic & p-value               \\ \hline
$\delta$  & 0.0681   & 0.00219         & 28          & 2.75$\times 10^{-9}$ \\
\end{tabular}
\end{table}

\subsection{Adjusting parameter $\lambda$.}

\begin{figure}
  \centering
  \includegraphics[width=\textwidth,height=\textheight,keepaspectratio]{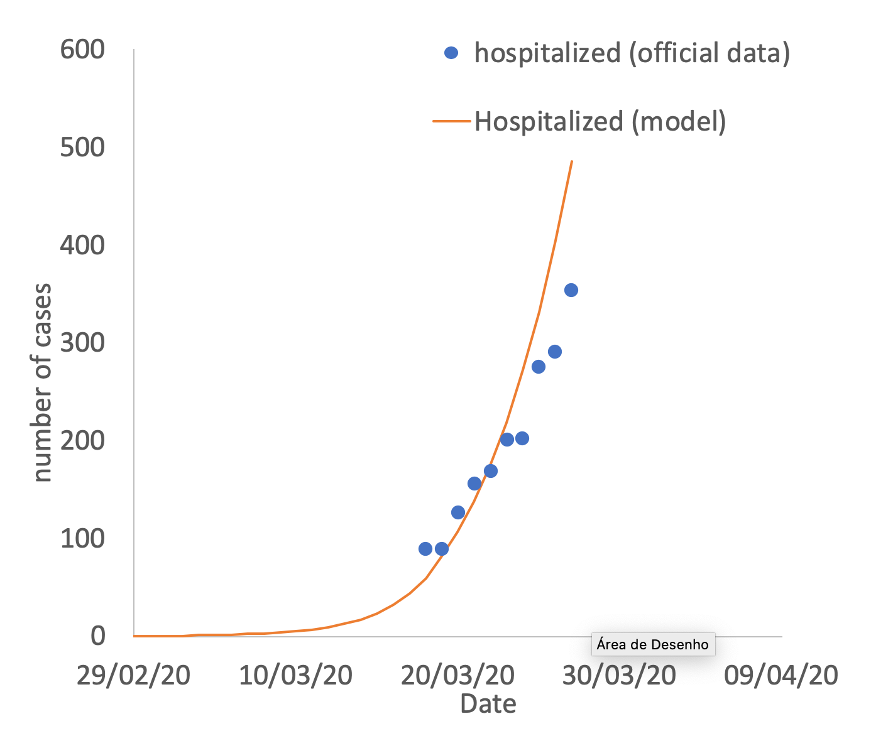}
  \caption{Graphical representation of the adjusted model to the hospitalized cases in Portugal.}
  \label{fig:fig3}
\end{figure}

Now, using the value for  $\delta$ found in 3.1, parameter $\lambda$ was obtained  with a new fit of Eq. 1 to the number of hospitalized cases in Portugal, which again was taken from \cite{16}. The best possible fit  gave an $R^2$ value of $\sim$0.98. The rest of the parameters are presented in table \ref{table3}, and shown in figure \ref{fig:fig3}.

\begin{table}
\caption{Table of fitted parameter  $\lambda$ as obtained using “NonLinearCurveFit” in Mathematica}
\label{table3}
\centering
\begin{tabular}{lllll}
\toprule
parameter & estimate & standard error & t-statistic & p-value               \\ \hline
$\lambda$ & 0.285   & 0.0141         & 20.07         & 2.98$\times 10^{-12}$ \\
\end{tabular}
\end{table}

\subsection{Fitting of parameter $\beta$ in period 0}

The parameters needed to fit Eq. 1 to the number of Portuguese active cases until the 18th of March, in order to determine the value of $\beta$, were obtained in 3.1, and 3.2. The number of Portuguese active cases was taken from \cite{16}. The fit was obtained with an $R^2$ of ~0.99. The other parameters for $\beta$ are given in table \ref{table4}. In this period, no measures were being taken so, $l=1$; and $d=0$.

\begin{table}
\caption{Table of fitted parameters  $\lambda$ and $\delta$ as obtained using “NonLinearCurveFit” in Mathematica}
\label{table4}
\centering
\begin{tabular}{lllll}
\toprule
parameter & estimate & standard error & t-statistic & p-value               \\ \hline
$\beta$ & 1.023   & 0.00263         & 388.7        & 5.19$\times 10^{-35}$ \\
\end{tabular}
\end{table}

\begin{figure}
  \centering
  \includegraphics[width=\textwidth,height=\textheight,keepaspectratio]{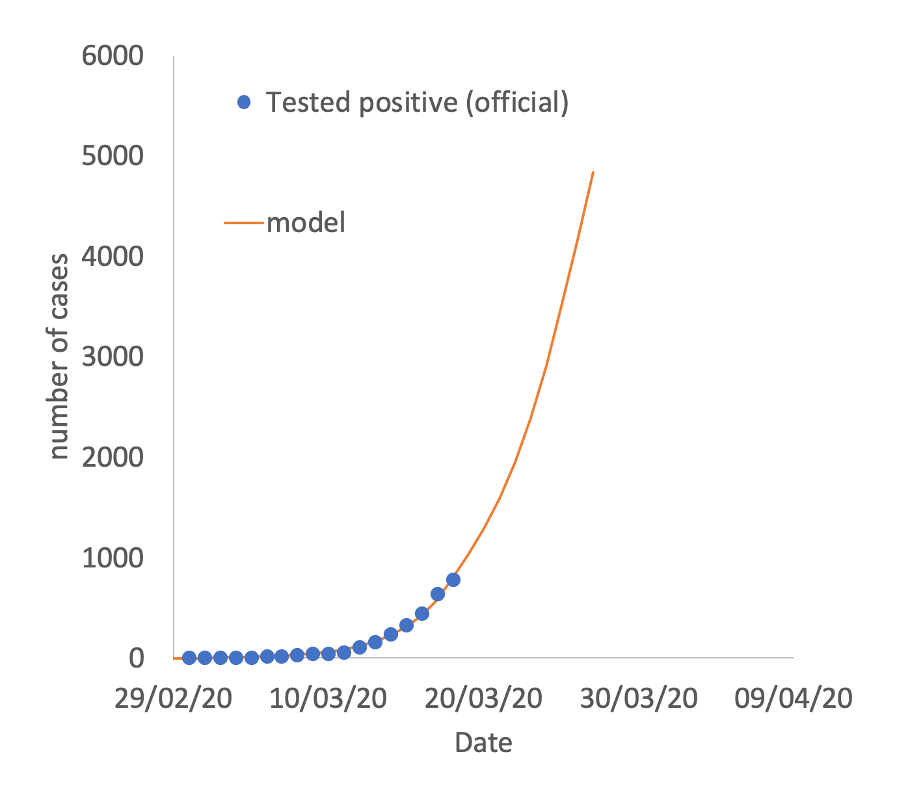}
  \caption{Graphical representation of the adjusted model to the Portuguese government official active cases.}
  \label{fig:fig4}
\end{figure}

A graphical depiction of the fitted model is given in figure \ref{fig:fig4}. 

\subsection{Discussion of the obtained values for $\lambda$, $\delta$, and $\beta$ for period 0}

According to the obtained values of  $\lambda$, $\delta$, and $\beta$, the mean time between onset of symptoms and hospitalization (for the 13\% of cases that need to be hospitalized) is of $\lambda^{-1}=3.51\pm 0.17$ days, and the mean time of those hospitalized until death is of $\delta^{-1}=14.6\pm 0.45$days. Unfortunately, I did not find any official data about these numbers in order to establish a comparison to ascertain their adequacy.

Also, the model assumes that all symptomatic and no asymptomatic cases are being tested, which may not reflect a realistic situation.  This is probably very conditional to the number of tests being performed by the country and may also vary with time.

\begin{table}
\caption{The obtained parameters of the model described in the previous sections}
\label{table5}
\centering
\begin{tabular}{ll}
\toprule
parameter      &  this work ($days^{-1}$)   \\ \hline
$\lambda$ & 0.285$\pm$0.00141 \\
$\delta$  & 0.0681$\pm$0.0002 \\
$\beta$   &1.023$\pm$0.00263   \\
\end{tabular}
\end{table}

\section{Time evolution of Sars-CoV-2 in Portugal, considering time adjustable parameters.}

In an epidemic outbreak such as the current one, governments and the population take protective measures to attempt to contain the spread of the virus. In this paper, parameters l and d, inspired by the paper from Xia \textit{et al}\cite{12,13} , are introduced as explained in section 3.  These parameters are meant to take into account two things:
\begin{itemize}
    \item Isolation and monitoring measures taken by the government (parameter $d$);
    \item Self-protection measures taken by the population (parameter $l$).
\end{itemize}

In a previous study, four different scenarios in addition to an “out-of-control” scenario, were devised. To note that we are currently working with $d_1=d_2=d_3=d_4=d$ and $l_1=l_2=l_3=l$.

The day considered for the initiation of these measures is day 33 corresponding to the 19th of March 2020 which is 5 days after the government implemented the measures, giving the virus another cycle of infection before the measures take effect. 

\subsection{Implementing a dynamic model to follow the trajectory of the curve.}

\begin{figure}
  \centering
  \includegraphics[width=\textwidth,height=\textheight,keepaspectratio]{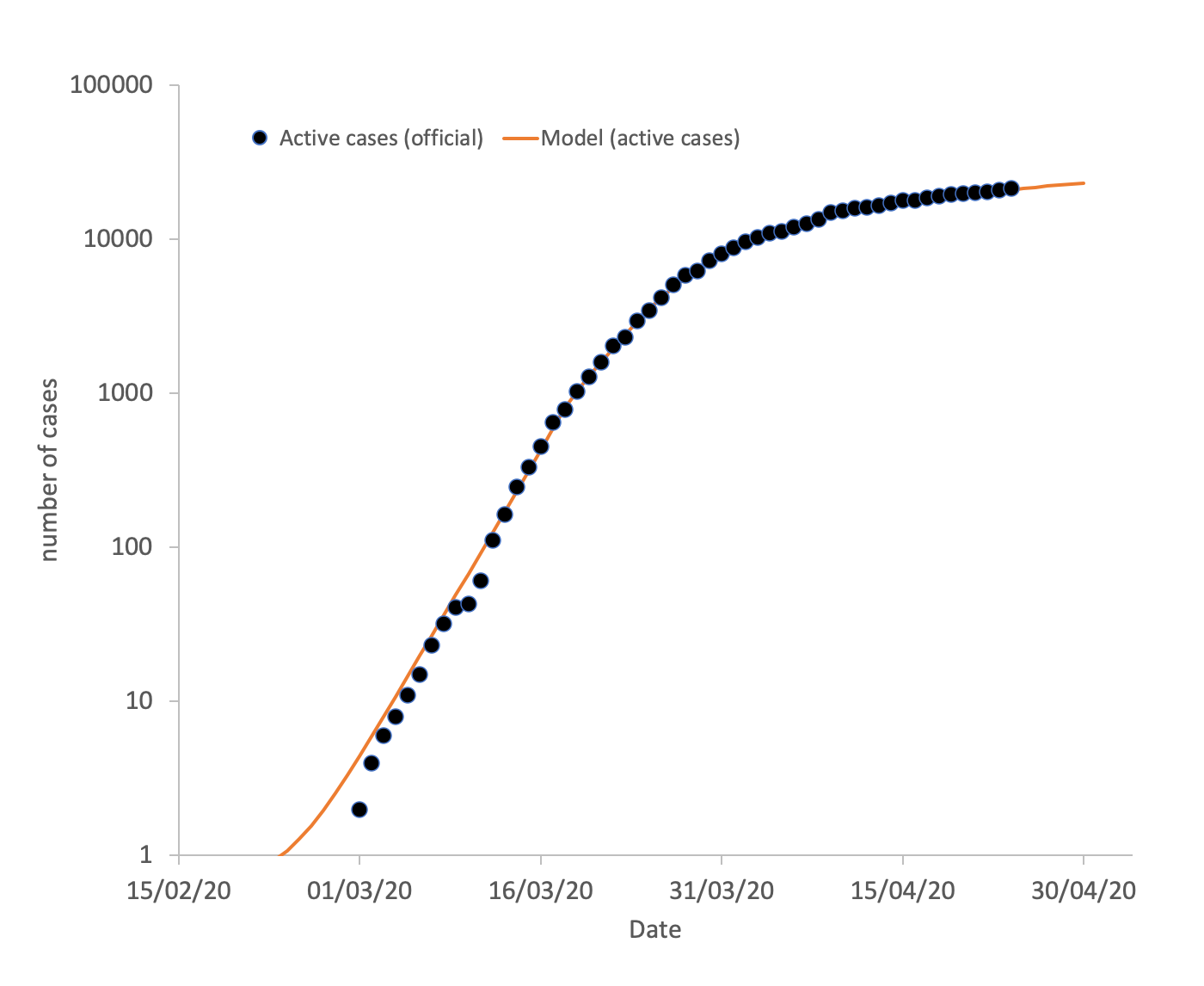}
  \caption{Graphical representation of official active cases in Portugal and the model. Blue dots represent official data, the orange line the model. }
  \label{fig:fig5}
\end{figure}

\begin{figure}
  \centering
  \includegraphics[width=\textwidth,height=\textheight,keepaspectratio]{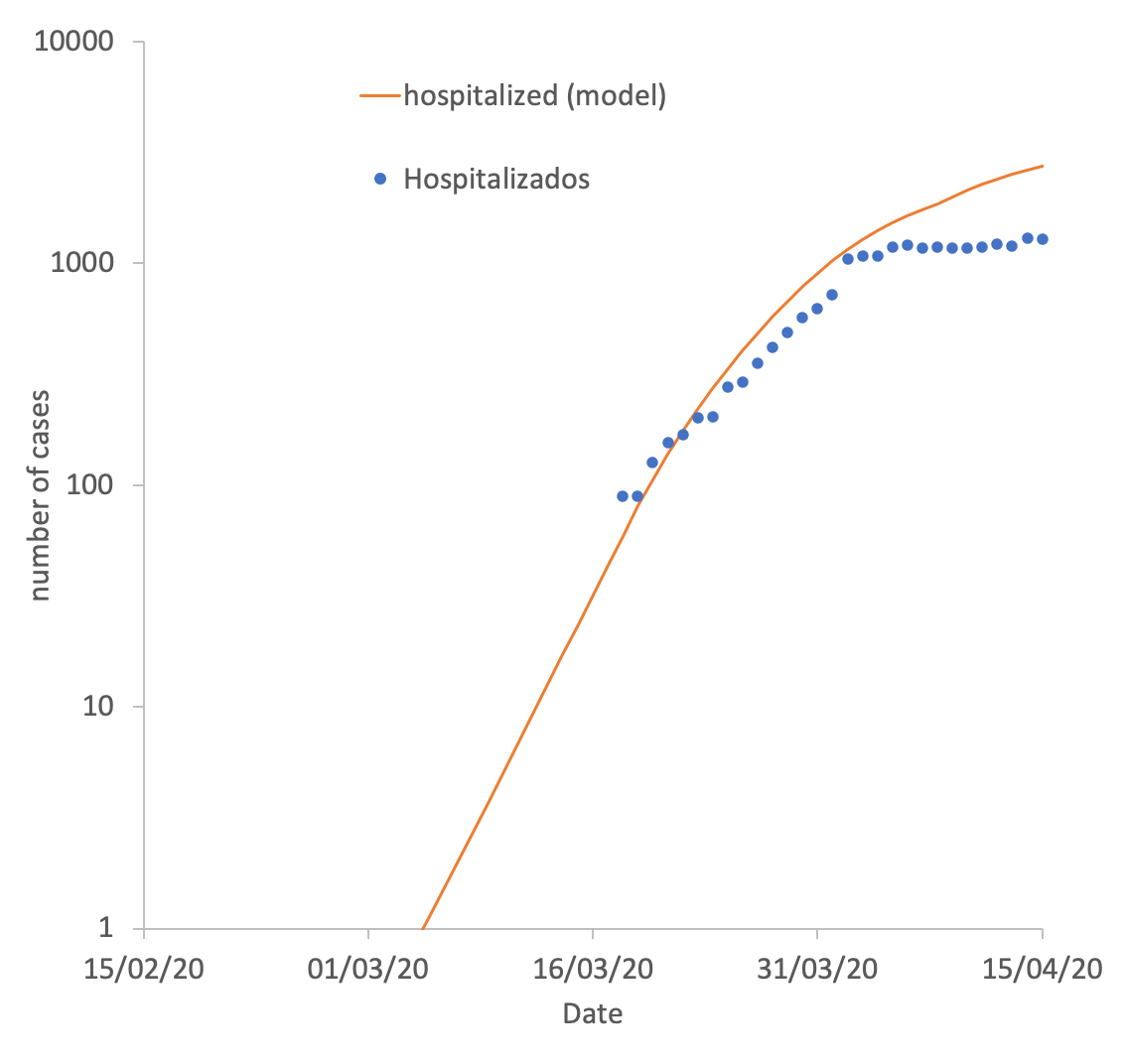}
  \caption{Graphical representation of hospitalized cases in Portugal. Blue dots represent official data, the orange line the model.}
  \label{fig:fig6}
\end{figure}

In the current analysis, instead of using set values from educated guesses as done in my previous work, the values of $\beta$, $d$ and $l$ are adjustable to the curve. In order to fit the results, four periods of five days each, from the 19th of March, were considered. In the first three periods, the same official data to fit the results was used. Initial conditions were taken from the previous period, and equation 1 was solved for the trajectory of the curve in the remainder of days in the same period. This procedure was performed for each period.  This allowed for a dynamic follow-up of the trajectory, adjusting the parameters to the best possible fit to the curve in each period.  The obtained results are summarized in table \ref{table6}. A graphical depiction of the evolution of the obtained curve for infected cases is given in \ref{fig:fig5}. For hospitalized cases, this is provided in figure \ref{fig:fig6}.

\begin{table}
\caption{Obtained parameters for each of the periods considered }
\label{table6}
\centering
\begin{tabular}{lllll}
\toprule
period   &  $\beta$ (days$^{-1}$) & $l$ (no units) & $d$ (days$^{-1}$)& $R^2$  \\ \hline
Period 1 (19th to 24th March 2020) & 1.101$\pm$0.0167 & 0.641$\pm$0.009 & 0.0353$\pm$.0816&0.999 \\
Period 2 (25th to 30th March 2020)  & 1.052$\pm$0.0315 & 0.379$\pm$0.011& 0.0100$\pm$0.105&0.999\\
Period 3 (31th March to 4th April 2020)  & 0.509$\pm$0.001& 0.489$\pm$0.009& 0.0100$\pm$0.033&0.999\\
Period 4 (5th to 10th April 2020)  & 0.571$\pm$0.008& 0.521$\pm$0.006& 0.0453$\pm$0.022&0.999\\
Period 5 (10th to 15th April 2020)  & 0.227$\pm$0.017& 0.569$\pm$0.043& 0.0100$\pm$0.046&0.999\\ 
\end{tabular}
\end{table}

As is clear from the figures, the number of active and hospitalized cases was  reached between the 10th and the 15th of April at around 1,250 cases, below the predictions of the model. In fact the number of hospitalized cases has reached a plateau. Current active cases are around $\sim$19,000. Considering that the number of ICU cases has been between 20-30\% of the total of hospitalized cases (see data), this would mean a number of about 200~300 cases in ICU. 

\begin{figure}[!htbp]
  \centering
  \includegraphics[width=\textwidth,height=\textheight,keepaspectratio]{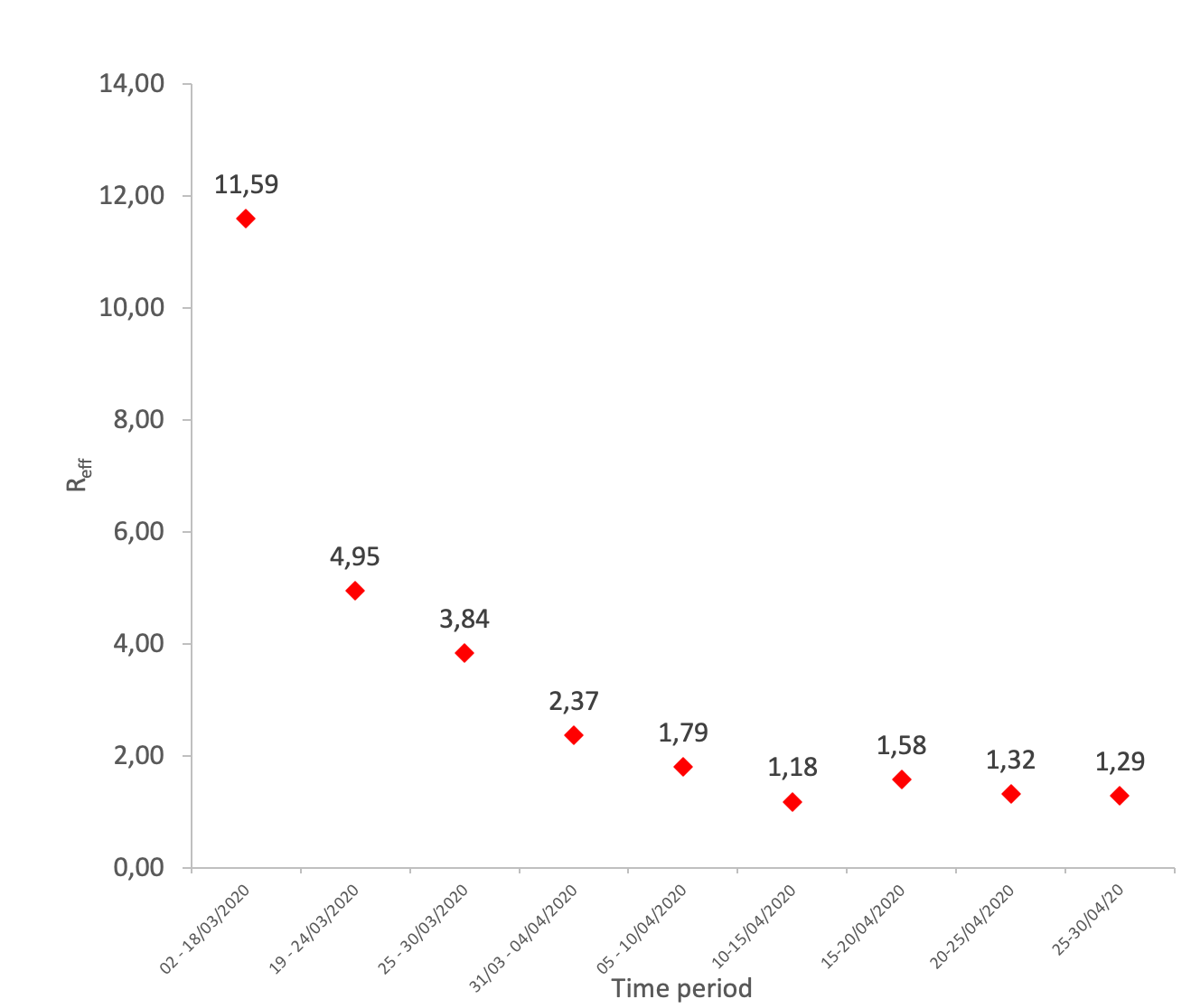}
  \caption{Graphical representation of $R_0$ for each time period.}
  \label{fig:fig7}
\end{figure}

The last period corresponds to a prediction of the values for the five days following period 5 (10th to the 15th of April 2020). Both $l$ and $\beta$ were extrapolated using a polinomial function, which enables the estimation of their value for the next period, as can be seen in table \ref{table8}:

\begin{table}
\caption{Obtained parameters for each of the periods considered }
\label{table8}
\centering
\begin{tabular}{llll}
\toprule
period   &  $\beta$ (days$^{-1}$) & $l$ (no units) & $\beta_{eff}$ (days$^{-1}$ \\ \hline
Period 0 (2nd to 19th March 2020) & 1.023$\pm$0.0026 & 1 & 1.023$\pm$0.0026 \\ \hline
Period 1 (19th to 24th March 2020) & 1.101$\pm$0.0167 & 0.641$\pm$0.009 & 0.706$\pm$.00146 \\ 
Period 2 (25th to 30th March 2020)  & 1.052$\pm$0.0315 & 0.379$\pm$0.011& 0.399$\pm$0.0166\\
Period 3 (31th to 4th April 2020)  & 0.509$\pm$0.001& 0.489$\pm$0.009& 0.249$\pm$0.005\\ 
Period 4 (5th to 10th April 2020) & 0.571$\pm$0.008 & 0.521$\pm$0.006 & 0.297$\pm$0.005 \\
Period 4 (10th to 15th April 2020) & 0.227$\pm$0.017 & 0.569$\pm$0.043 & 0.129$\pm$0.014 \\
\hline
Period 6 (15th to 20th April 2020)  & 0.301 & 0.450 & 0.136 \\
\end{tabular}
\end{table}

In this table, we are already considering the effective value of the transmission rate, $\beta_{eff}(t)$:
\begin{equation}
\beta_{eff}(t)=l\cdot \beta(t)
\end{equation}

\subsection{Estimating $R_0$ for this model.}

The obtained values for $\beta$, allow for an estimate of the basic reproduction number for all considered periods.

Following the formalism proposed by Van der Driessche and Watmough\cite{van}, and $\alpha=0.13$ we can define $\mathscr{F}$ and $\mathscr{V}$,  that the equilibrium state is given by $(S_0,0,0,0)$:

\begin{equation}
\mathscr{F} = 
\begin{pmatrix}
 l\frac{S\beta}{N}(A+I+H)
 \\
0 \\
0 \\
0 \\
\end{pmatrix}
\quad
\textrm{and}
\quad
\mathscr{V} = 
\begin{pmatrix}
(\sigma+d)E\\
-(1-\gamma)\sigma E+(k+d)A \\
-\gamma\sigma E+\alpha\lambda I+(1-\alpha)(k+d)I \\
-\alpha\lambda I+kH+(\delta+d)H \\
\end{pmatrix}
\end{equation}

We can then determine the Jacobian matrices:

\begin{equation}
F = 
\begin{pmatrix}
0 &  l\frac{S\beta}{N} & l\frac{S\beta}{N} & l\frac{S\beta}{N}\\
0 &0 &0 &0 \\
0 &0 &0 &0 \\
0 &0 &0 &0  \\
\end{pmatrix}
\quad
\textrm{and}
\quad
V = 
\begin{pmatrix}
\sigma + d &  0& 0 & 0\\
-(1-\gamma)\sigma & k+d &0 &0 \\
-\gamma\sigma &0 & \scriptstyle \alpha\lambda + (1-\alpha)(k+d) &0 \\
0 &0 & -\alpha\lambda & \scriptstyle k+(\delta+d)  \\
\end{pmatrix}
\end{equation}

We also need $V^{-1}$:
\begin{equation}
V^{-1} = 
\begin{pmatrix}
\frac{1}{\sigma + d} &  0& 0 & 0\\
\frac{(1-\gamma)\sigma}{(\sigma + d)(k+d)} & \frac{1}{k+d} &0 &0 \\
\scriptstyle \frac{\gamma\sigma}{(\alpha\lambda + (1-\alpha)(k+d))(\sigma + d)} &0 & \scriptstyle \frac{1}{\alpha\lambda + (1-\alpha)(k+d)} &0 \\
\frac{\alpha\lambda\gamma\sigma}{(k+(\delta+d))(\alpha\lambda + (1-\alpha)(k+d))(\sigma + d)} &0 &\frac{\alpha\lambda}{k+(\delta+d)}  & \scriptstyle -\frac{1}{k+(\delta+d)}  \\
\end{pmatrix}
\end{equation}

And $R_c$ will simply be 
\begin{equation}
R_c=\rho(FV^{-1})=l\frac{S\beta}{N}\left( \frac{(1-\gamma)\sigma}{(\sigma + d)(k+d)}+\frac{\gamma\sigma}{(\alpha\lambda + (1-\alpha)(k+d))(\sigma + d)}\right)+\frac{\alpha\lambda\gamma\sigma}{(k+(\delta+d))(\alpha\lambda + (1-\alpha)(k+d))(\sigma + d)}
\end{equation}

For $R_0$, $l=1$, and $d=0$, and so:

\begin{equation}
R_0=\rho(FV^{-1})=\frac{S_0\beta}{N}\left( \frac{1-\gamma}k+\frac{\gamma}{\alpha\lambda + (1-\alpha)k}+\frac{\alpha\lambda\gamma}{(k+\delta)(\alpha\lambda + (1-\alpha)k)}\right)
\end{equation}

This gives a value of $R_0=11.77$, indicating a very active spread of the virus in this first stage, although this value seems higher than most studies. In this model, given that cases will only resolve after a period of about $ k^{-1}\sim10$ days, during which time they will be able to infect another individual. Also, there is a very high uncertainty for the appropriate incubation time of this virus, which some authors claim can be of up to 27 days \cite{22,23}. This is still something that needs to be analysed in more detail.

This permits us to estimate the value for the evolution of the basic reproduction number $R_c$ with time, as can be seen in figure \ref{fig:fig7}.

As can be seen the value of $R_0$ is diminishing with time, with a predicted value already very close to $1$ for the period of the 20th to the 25th of April 2020, although the value has clearly increased from a minimum in the week from the 10th to the 15th of April.

\section{Conclusions}

SEIR models, although sometimes blunt in their deterministic approach, are valuable tools to model epidemic outbreaks. If used carefully and dynamically (i.e. with adjustable parameters on given periods) it is possible to accurately predict the trajectory of an epidemic curve within a five to ten-day period. Protective and isolation measures are currently the only weapon at our disposal to control this epidemic. SEIR models are effective in taking these measures into account by the insertion of one or two extra parameters. Variation of such parameters can be demonstrative of the power of any given measure, be it in terms of self-protection (washing hands frequently, social isolation, etc.) or community measures (closing down schools, parks, etc.).

\section{Limitations of this study and scope of application}

This study is intended for academic purposes only and should not be used in any other way. It intends to provide tools for scientists to model the trajectory of active and hospitalized cases in the current pandemic. 
With this model, and although in earlier predictions the values were of $\sim$13,000 active cases and $\sim$2,500 hospitalized cases, it seems now that a maximum of $\sim$19,000 active cases and $\sim$1,250 hospitalized cases will or has already been reached in this period. The value of $R_{eff}$ is now approaching one which means that in the next few days we will see the number of active cases starting to finally diminish.
This model assumes that only symptomatic cases are being tested. Given that the number of tests in Portugal is now one of the highest in the World, it is very likely that non priority cases are also being tested now, which means that the model is starting to fail, as asymptomatic cases begin being tested. This could explain the plateau reached in the number of hospitalized cases, which, as we know represent 13\% of the total number of symptomatic cases. If this hypothesis is proven to be correct, the country may have already reached the peak of active cases (both symptomatic and asymptomatic), but is simply testing asymptomatics now leading to an increase in the number of confirmed cases, but a constant number of hospitalized and ICU cases.

\bibliographystyle{unsrt}  


\begin{thebibliography}{1}

\bibitem{1}Novel Coronavirus (2019-nCoV) situation reports - World Health Organization (WHO) https://www.who.int/emergencies/diseases/novel-coronavirus-
2019/situation-reports/ accessed 6th April 2020
\bibitem{2}Worldometer (https://www.worldometers.info/coronavirus/) accessed 6th April 2020
\bibitem{3}	Coronavirus Disease (COVID-19) – Research and Statistics, Max Roser, Hannah Ritchie and Esteban Ortiz-Ospina (2020), ourworldindata.org/coronavirus, accessed 6th April 2020
\bibitem{4}	The species Severe acute respiratory syndrome-related coronavirus: classifying 2019-nCoV and naming it SARS-CoV-2, Gobalenya AE, Baker SC, Baric RS, et al. (March 2020). Nature Microbiology: 1–9. doi:10.1038/s41564-020-0695-z. PMID 32123347.
\bibitem{5}Epidemiological and clinical characteristics of 99 cases of 2019 novel coronavirus pneumonia in Wuhan, China: a descriptive study Chen N, Zhou M, Dong X, et al (February 2020). Lancet. 395 (10223):
507–13. doi:10.1016/S0140-6736(20)30211-7. PMID 32007143
\bibitem{6}	COVID-19: what is next for public health?. Heymann DL, Shindo N, et al. (WHO Scientific and Technical Advisory Group for Infectious Hazards) (February 2020). Lancet. Elsevier BV. 395 (10224): 542–545. doi:10.1016/s01406736(20)30374-3. PMID 32061313.
\bibitem{7}Case-fatality estimates for COVID-19 calculated by using a lag time for fatality Wilson N, Kvalsvig A, Telfar Barnard L, Baker MG. Emerg Infect Dis. 2020 Jun [date cited].  https://doi.org/10.3201/eid2606.200320
\bibitem{8}Case-Fatality Rate and Characteristics of Patients Dying in Relation to COVID-19 in Italy Onder G, Rezza G, Brusaferro S.. JAMA. Published online March 23, 2020. doi:10.1001/jama.2020.4683
\bibitem{9}The Epidemiological Characteristics of an Outbreak of 2019 Novel Coronavirus Diseases (COVID-19) in China. Zhonghua Liu, Xing Bing, Xue Za Zhi. 2020;41(2):145–151. DOI:10.3760/cma.j.issn.0254-6450.2020.02.003.
\bibitem{10}	WHO Director-General’s opening remarks at the media briefing on COVID-19 - 13 March 2020 World Health Organization
https://www.who.int/dg/speeches/detail/who-director-general-s-openingremarks-at-the-mission-briefing-on-covid-19—13-march-2020 accessed 18 March 2020
\bibitem{11}	The Extraordinary Decisions Facing Italian Doctors, Yascha Mounk, The Atlantic, 11 March 2020
https://www.theatlantic.com/ideas/archive/2020/03/who-gets-hospitalbed/607807/ accessed 18 March 2020
\bibitem{12}	Modeling the Transmission of Middle East Respirator Syndrome Corona Virus in the Republic of Korea Xia Z-Q, Zhang J, Xue Y-K, Sun G-Q, Jin Z(2015). PLoS ONE 10(12): e0144778. doi:10.1371/ journal.pone.0144778
\bibitem{13}	Predicting the evolution Of Sars-CoV-2 in Portugal using an adapted SIR Model previously used in South Korea for the MERS outbreak P Teles - arXiv preprint arXiv:2003.10047, 2020

\bibitem{14}	A Time-dependent SIR model for COVID-19, Yi-Cheng Chen, Ping-En Lu, Cheng-Shang Chang arXiv:2003.00122v1 (28 Feb 2020)

\bibitem{15}Planning as Inference in Epidemiological Models rank Wood, Andrew Warrington, Saeid Naderiparizi, Christian Weilbach, Vaden Masrani, William Harvey, Adam Scibior, Boyan Beronov, Ali Nasseri

\bibitem{16}	Ponto da Situação atual em Portugal,	Direcção Geral de Saúde
(in Portuguese), https://covid19.min-saude.pt/ponto-de-situacao-atual-emportugal/ accessed 5th April 2020
\bibitem{17}	The incubation period of coronavirus disease 2019 (COVID-19) from publicly reported confirmed cases: Estimation and application. Lauer SA et al. Ann Intern Med 2020 Mar 10; [e-pub]. (https://doi.org/10.7326/M20-0504)
\bibitem{18}	Dipartimento della Protezione Civile COVID-19 Italia - Monitoraggio della situazione (in Italian) http://tiny.cc/zv3slz accessed 18 March 2020
\bibitem{19}	Epidemiology and Transmission of COVID-19 in Shenzhen China: Analysis of 391 cases and 1,286 of their close contacts Qifang Bi, Yongsheng Wu, Shujiang Mei, Chenfei Ye, Xuan Zou, Zhen Zhang, Xiaojian Liu, Lan Wei, Shaun A Truelove, Tong Zhang, Wei Gao, Cong Cheng, Xiujuan Tang, Xiaoliang Wu, Yu Wu, Binbin Sun, Suli Huang, Yu Sun, Juncen Zhang, Ting Ma, Justin Lessler, Teijian Feng medRxiv 2020.03.03.20028423; doi: https://doi.org/10.1101/2020.03.03.20028423
\bibitem{20}	Wolfram Research, Inc., Mathematica, Version 12.1, Champaign, IL (2020).
\bibitem{21}	NonLinearModelFit http://tiny.cc/pw3slz accessed 18 March 2020
\bibitem{van} Driessche, P. van den and J A Watmough. “Reproduction numbers and sub-threshold endemic equilibria for compartmental models of disease transmission.” Mathematical biosciences 180 (2002): 29-48 .
\bibitem{22}	Coronavirus incubation could be as long as 27 days, Chinese provincial government says - Reuters, Feb. 22, 2020
\bibitem{23}	Presumed Asymptomatic Carrier Transmission of COVID-19. JAMA. Bai Y, Yao L, Wei T, et al., JAMA. Published online February 21, 2020. doi:10.1001/jama.2020.2565



\end{thebibliography}

\end{document}